\documentclass[a4paper,11pt]{article}
\usepackage[utf8]{inputenc}

\usepackage{amsfonts}
\usepackage{amssymb}
\usepackage{graphicx}
\usepackage{amsmath}
\usepackage{enumerate}
\usepackage{color}
\usepackage{cite}
 \usepackage{multirow}
\usepackage{float}

\RequirePackage{mathrsfs} \RequirePackage[sc]{mathpazo}
\RequirePackage{wasysym} \RequirePackage{setspace}\textheight=650pt
\title{ {\bf \Large 
More Insight into  Microscopic Properties of RN-AdS Black Hole Surrounded by Quintessence via an  Alternative Extended Phase Space.}}
\author{M. Chabab$^{1}$\footnote{mchabab@uca.ac.ma (Corresponding author)}, H. El Moumni$^{1,2}$\footnote{hasanelm@yahoo.fr}, S. Iraoui$^{1}$\footnote{s.iraoui@edu.uca.ma}, K. Masmar$^{1}$\footnote{karima.masmar@edu.uca.ac.ma}, S. Zhizeh$^1$\footnote{sara.zhizeh@edu.uca.ac.ma}\\
	\\ 
	{\small $^{1}$ High Energy and Astrophysics Laboratory, Physics Department, FSSM, 
	}\\
         {\small Cadi Ayyad University, P.O.B. 2390 Marrakech, Morocco.
	}\\
	{\small $^{2}$ LMTI, Physics Department, Faculty of Sciences,  Ibn Zohr University, Agadir, Morocco. }
}

\begin{document}
\maketitle
\begin{abstract}
{\noindent}
In this work we study   the phase transition of the charged-AdS black hole surrounded by quintessence via an alternative extended phase space defined by the charge square $Q^2$ and her conjugate  $\Psi$, a quantity proportional to the  inverse of horizon radius,  while the cosmological constant is kept fixed. The equation of state is derived under the form $Q^2=Q^2(T,\Psi)$ and the critical behavior of such black hole analyzed. In addition, we explore the connection between  the microscopic structure and Ruppeiner geothermodynamics. We also find that, at certain points of the phase space, the Ruppeiner curvature is characterized by the presence of singularities that are interpreted as phase transitions.

\end{abstract}

\newpage
\section{Introduction}

 Since  the discovery of the phase transition by Hawking and Page \cite{2},  an important line of research has been devoted to the
 study the black holes thermodynamics, with the aim to shed more light on our understanding of quantum gravity.
Recently, several papers have considered the cosmological constant $\Lambda$ as a dynamical variable \cite{12, 13}, while  others  have treated $\Lambda$ as a
thermodynamic variable \cite{14}, identified to the thermodynamical pressure. Using the latter proposal, many efforts have been devoted to disclose the phases transitions of black holes in AdS space \cite{KM,our} and to reinforce the analogy between the critical  behaviors of the Van der Waals gas and the charged AdS black hole \cite {our1,our2,our3,our4,our5,our6,our7,ElMoumni:2016eqh}.

On the other hand, the cosmological constant is a serious candidate for explaining  the puzzle of dark energy. This energy, with a negative pressure, 
is responsible of the observed accelerating expansion of our Universe, considered as one of the most  fascinating results of observational cosmology. 
In addition to the dark energy, others viable candidates might play a role in such expansion such as quintessence, phantom and quantom. 

In this work, we will only deal with quintessence's existence and its effects on the black hole spacetime configuration.
Since the first paper on black holes surrounded by quintessence \cite{zh38}, this area of physics  becomes  a fertile ground to probe
the impact of quintessence dark energy on various features of the black holes, as quasi-normal modes \cite{24,25},
thermodynamical properties \cite{26,27,28}, P-V criticality in extended phase space and heat engine \cite{29,machine}, the stability analysis \cite{ma}, as well as  the  phase transitions in a holographic framework \cite{holo}.

Our aim in  this work is to study the critical behavior and the geometrical thermodynamics 
of charged-AdS black hole surrounded by quintessence in an alternative extended phase  space, and then establish a link with the microscopic structure.  Compared to the usual approach, here we consider the cosmological constant as a fixed parameter 
while the square of the charge $Q^2$ is treated as a thermodynamical variable. Then, the occurence of the phase transition is analysez in $(Q^2,\Psi)$-diagram, 
where  $\Psi$ is the conjugate quantity to $Q^2$.

The  paper is organized as follow: In the next section, we briefly introduce the main ingredients to be used by the new alternative extended phase space.  The new form of the first law of the thermodynamics as well as the Smarr formula in this black hole background are presented.  Then we proceed
with the calculations of various thermodynamic quantities 
and show the small-large black hole phase transition. The coordinates of the critical point and the critical exponents are also derived, in addition to the coexistence line.
In section 3,  we reveal the microscopic properties of charged AdS black holes surrounded by quintessence via the thermodynamic geometry. More specifically,  we evaluate the Ricci scalar curvature of the Ruppeiner metric and  look for more  insights into the nature of interactions among the black constituents. The last section is devoted to our conclusion.

\section{Thermodynamics of charged black hole surrounded by quintessence in $(Q^2,\Psi)$-plan}
We start by writing the metric of the spherically symmetric  charged-AdS black hole surrounded by quintessence  as \cite{zh38}
\begin{equation}
f(r)=1-\frac{2 M}{r}+\frac{Q^2}{r^2}+\frac{\Lambda  r^2}{3}-\frac{a}{r^{3\omega+1}},
\end{equation}

where $M$ and $Q$ are the ADM mass and charge of the black hole respectively,
$\omega_q$ represents the state parameter while $a$ reads as  the normalization factor related to the density of quintessence.
The state parameter of  quintessence dark energy is restricted to lie in $ - 1 < \omega_q < - 1 / 3$, while $\omega_q < - 1$ in  the case of phantom
dark energy. The normalization factor  $a$ being always positive, the density
of quintessence can be expressed by the formula,

\begin{equation}
 \rho_q=-\frac{a}{2}\frac{3 \omega_q}{r^{3(\omega_q +1)}}.
\end{equation}

The black hole mass $M$ can be expressed in term of the horizon radius $r_+$ as

\begin{equation}
M=\frac{1}{6} \left(-3 a r_{+}^{-3 \omega_q }+\frac{3 Q^2}{r_{+}}-\Lambda  r_{+}^3+3 r_{+}\right).
\end{equation}

while Hawking temperature and the entropy read as,

\begin{equation}\label{temp}
T=\left.\frac{f(r)}{4\pi }\right|_{r=r_+}=\frac{r_{+} \left(3 a \omega_q  r_{+}^{-3 \omega_q }+\frac{3 r_{+}^3}{l^2}+r_{+}\right)-Q^2}{4 \pi  r_{+}^3},
\end{equation}

\begin{equation}
	S=\int_{0}^{r_+} \frac{1}{T} \left(\frac{\partial M}{\partial r_+}\right)dr_{+}=\pi r_+^2.
	\end{equation}

Now, we probe the thermodynamical proprieties of a such black hole in the new extended phase space. The latter is made by considering the entropy $S$, the square of the charge $Q^2$, the normalisation factor $a$ and the  cosmological constant as independent parameter. In our scenario the analogy between the pressure and the cosmological constant still holds  via the usual formula
\begin{equation}
 P=-\frac{\Lambda}{8\pi}.
\end{equation}

Under these considerations, the mass of the black hole, identified to the enthalpy, is then given by,
\begin{equation}
 M(S,Q^2,P,a)=\frac{1}{6} \left(\frac{S (8 P S+3)+3 \pi  Q^2}{\sqrt{\pi } \sqrt{S}}-3 a \pi ^{3 \omega_q /2} S^{-3 \omega_q /2}\right).
\end{equation}

The intensive parameters respectively conjugate to S, $Q^2$, $P$ and $a$ are  defined  by
\begin{equation}
 T \equiv \left.\frac{\partial M}{\partial S}\right|_{P,Q^2,a},\quad \Psi\equiv \left.\frac{\partial M}{\partial Q^2}\right|_{S,P,a},\quad  V\equiv \left.\frac{\partial M}{\partial P}\right|_{S,Q^2,a}
\;\; and \;\;\; \mathcal{A}\equiv \left.\frac{\partial M}{\partial a}\right|_{S,Q^2,P},
\end{equation}

where $T$ denotes the temperature, the  new quantity $\Psi$ is the inverse of the specific volume $\Psi=\frac{1}{v}$ with  $v=2r_+$ in the natural unit. The thermodynamical volume is obtained via  
$V=\int 4 S dr_+=4 \pi r_+^3 /3$ while the quantity conjugate to the factor $a$ is $\mathcal{A}= -\frac{1}{2} r_{+}^{-3 \omega_q }$.  In this context the  new  first law
of black hole thermodynamics in this alternative extended phase space is written as, 

\begin{equation}
 d M = T dS + \Psi dQ^2+ V dP+ \mathcal{A} da.
\end{equation}
and the Smarr relation reads as
\begin{equation}
 M= 2 T S + \Psi Q^2 -2 V P+ (1+3\omega_q)\mathcal{A}a.
\end{equation}

It is worth to notice that, in the first law of the thermodynamics,  the usual term $\Phi dQ$, where $\Phi$ denotes the electric potential, has been changed to $\Psi dQ^2$ in our alternative phase space.   This modification leads to a novel behavior not present in the standard phase space associated with $dM = S dT +  \Phi dq$.
The small-large black hole phase transition occuring  in (P,v)-plan,  also shows up  in $ (Q^2,\Psi)-$ diagram. In the subsequent, we will study this phase transition: More precisely, we will derive the  coordinates of the critical point, study the Gibbs free energy as well as the critical exponents. 

By inverting Eq.\ref{temp}, we obtain the equation of state $Q^2(T,\Psi)$ as 

\begin{equation}\label{state}
Q^2=r_{+} \left(3 a \omega_q  r_{+}^{-3 \omega_q }+\frac{3 r_{+}^3}{l^2}-4 \pi  r_{+}^2 T+r_{+}\right).
\end{equation}

This equation is depicted in Figure \ref{fig1} for different values of temperature and normalisation factor $a$. 

\begin{center}
\begin{figure}[!ht]
\centering
\begin{tabbing}
\hspace{9cm}\=\kill
\includegraphics[scale=.9]{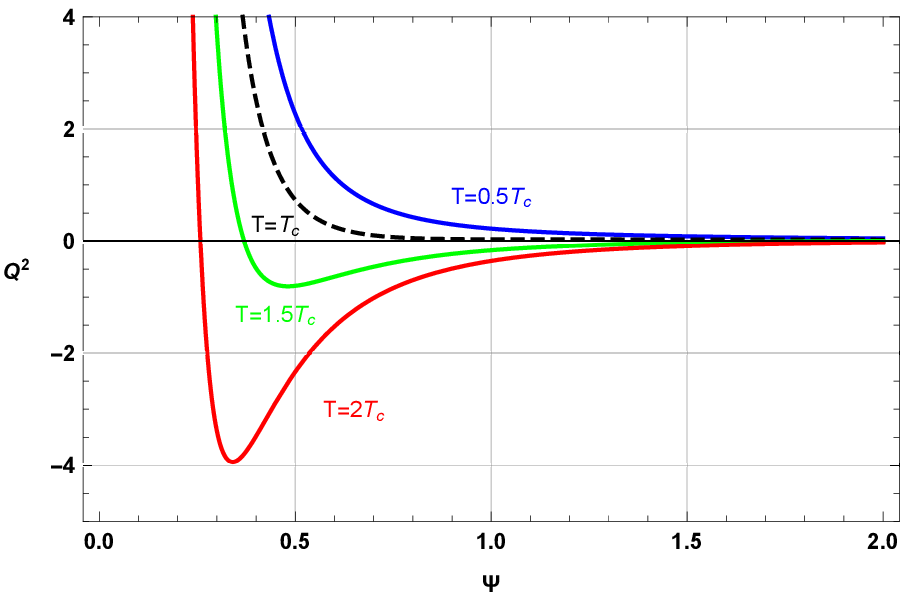}\> \includegraphics[scale=.9]{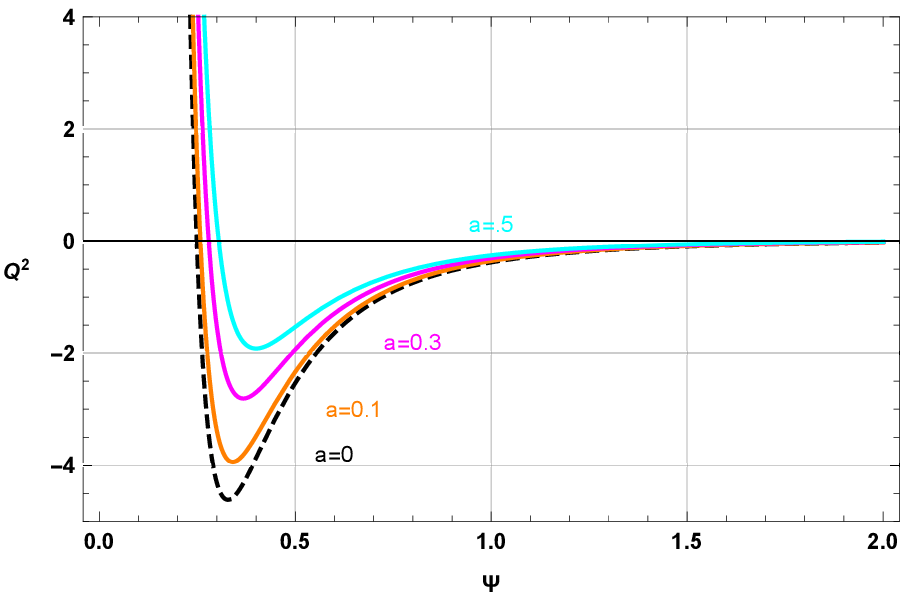} 
\end{tabbing}
\vspace*{-.2cm} \caption{\footnotesize {\bf Left:} The behavior of the isothermal line in $Q^2-\Psi$ diagram of the of charged black hole surrounded by quintessence, we have set $l=1$ and  $a=.1$  
{\bf Right:} The effect of the parameter $a$ on the isothermal line up to the critical temperature. }\label{fig1}
\label{fig1}
\end{figure}
\end{center}

Fig. \ref{fig1} shows a region with negative $Q^2$ having no physical meaning. This region also occurs  in the case of Van der Waals gas, where the pressure becomes negative for some values of temperature.
The oscillating part of the isotherm line corresponds to the instability region with $\left.\frac{\partial Q^2}{\partial \Psi}\right|_T >0 $. However, this instability can be removed through the Maxwell equal area construction \cite{Spallucci:2013jja}.
\begin{equation}
 \oint \Psi dQ^2=0.
\end{equation}

Evidently, 
 for $T > T_c$
there is an inflection
point and the behavior is reminiscent of the Van der Waals one.
The coordinates of the critical points are given by the roots of the following system,

\begin{equation}
 \left.\frac{\partial Q^2}{\partial \Psi}\right|_{T_c}=0,\quad \left.\frac{\partial^2 Q^2}{\partial \Psi^2}\right|_{T_c}=0.
\end{equation}

Thus one finds,
\begin{equation}
 T_c=\frac{2 \sqrt{6}-3 a l}{6 \pi  l},\;\;\; Q^2_c=\frac{l^2}{36},\;\;\; \Psi_c=\sqrt{\frac{3}{2l^2}}.
\end{equation}

\begin{center}
\begin{figure}[!ht]
\centering
\begin{tabbing}
\hspace{9cm}\=\kill
\includegraphics[scale=.85]{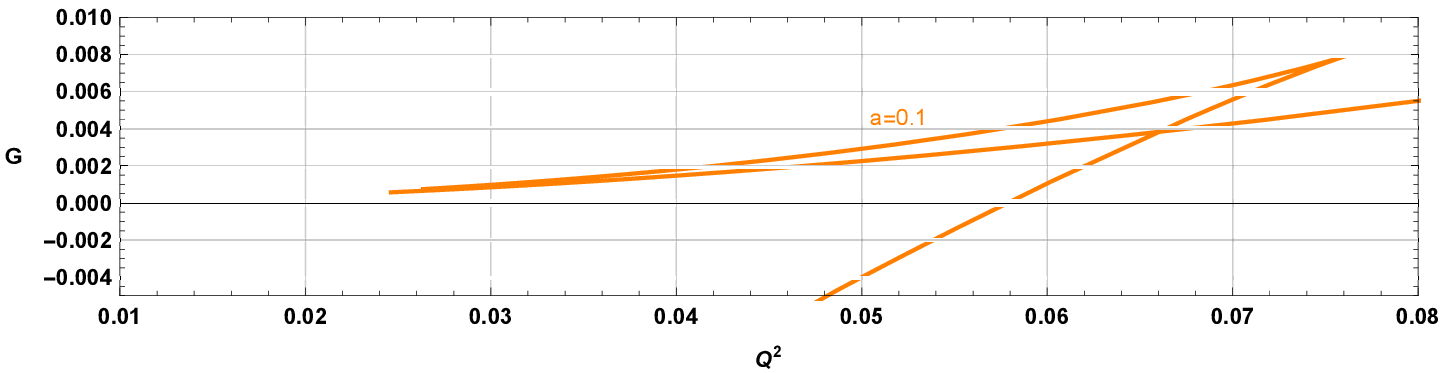}\\\includegraphics[scale=.85]{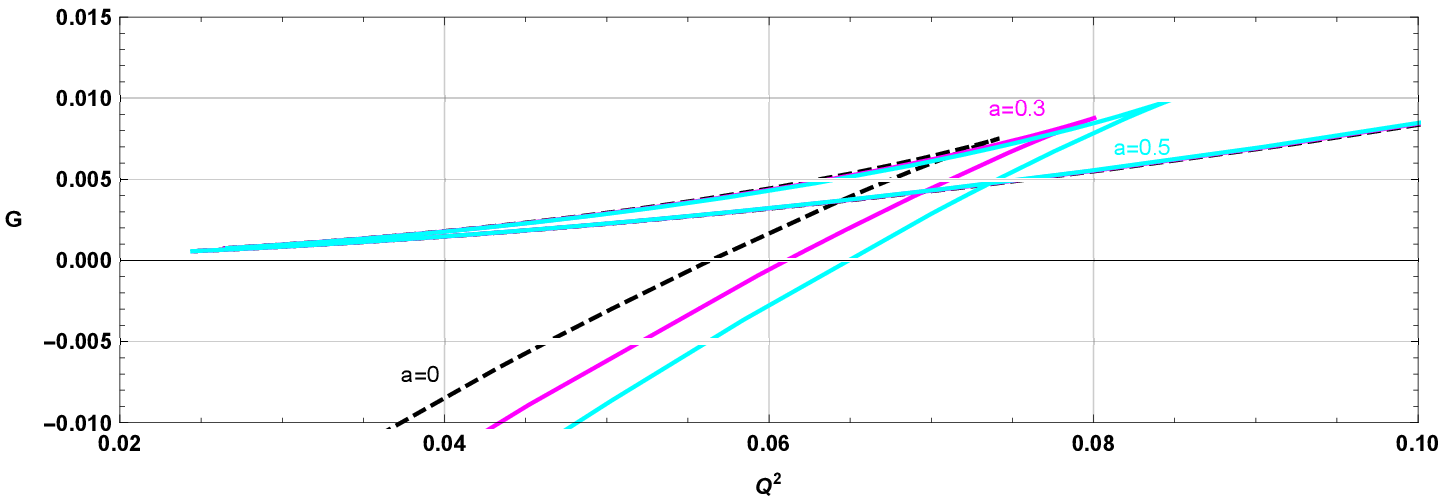} 
\end{tabbing}
\vspace*{-.2cm} \caption{\footnotesize The Gibbs free energy $G$ of AdS black hole surrounded  by quintessence versus the charge square $Q^2$ for a temperature larger than the critical one and different values of the normalisation factor $a$ }\label{fig2}
\end{figure}
\end{center}

As a byproduct, the resulting universal number which depends on the parameter $a$ and the AdS radius $l$ is, 

\begin{equation}
 \rho_c=Q^2_c T_c \Psi_c=\frac{4-\sqrt{6} a l}{144 \pi },
\end{equation}
If we set $a=0$, the usual results presented in \cite{micro} is recovered.
Furthermore, to  proceed further with the transition de phase in the thermal picture,  we analyze the Gibbs free energy given by,
\begin{equation}
G(T,Q^2,a)=M-T S=\frac{4 l^2 \Psi ^2 \left(2 \Psi  \left(6 Q^2 \Psi -a 8^{\omega_{q} } (3 \omega_{q} +2)
	\Psi ^{3 \omega_{q} }\right)+1\right)-1}{32 l^2 \Psi ^3},\end{equation}

and plot in Fig.\ref{fig2} $G$ as a function of the square of charge $Q^2$ in the scenario where $T> T_c$ for different values of the normalisation factor $a$.

\begin{center}
\begin{figure}[!ht]
\centering
\begin{tabbing}
\hspace{9cm}\=\kill
\includegraphics[scale=.8]{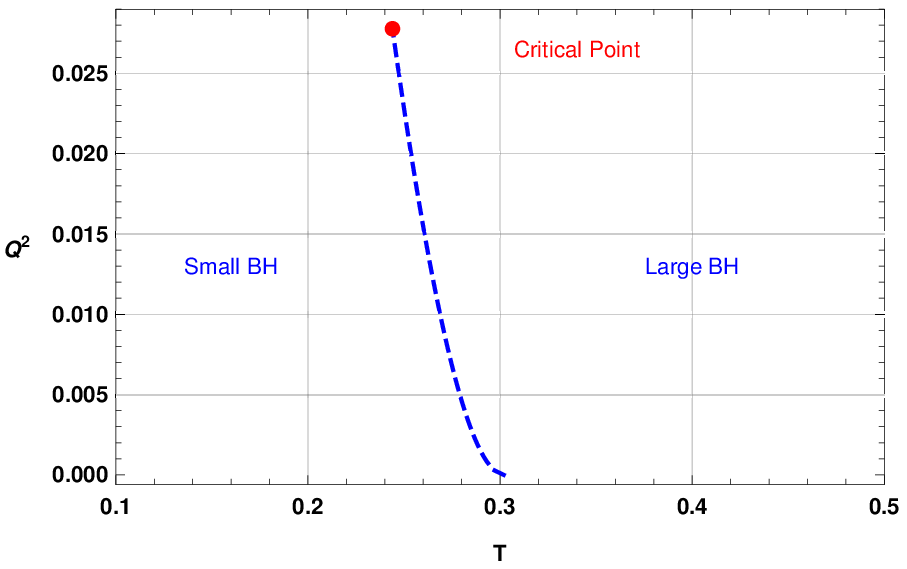}\> \includegraphics[scale=.8]{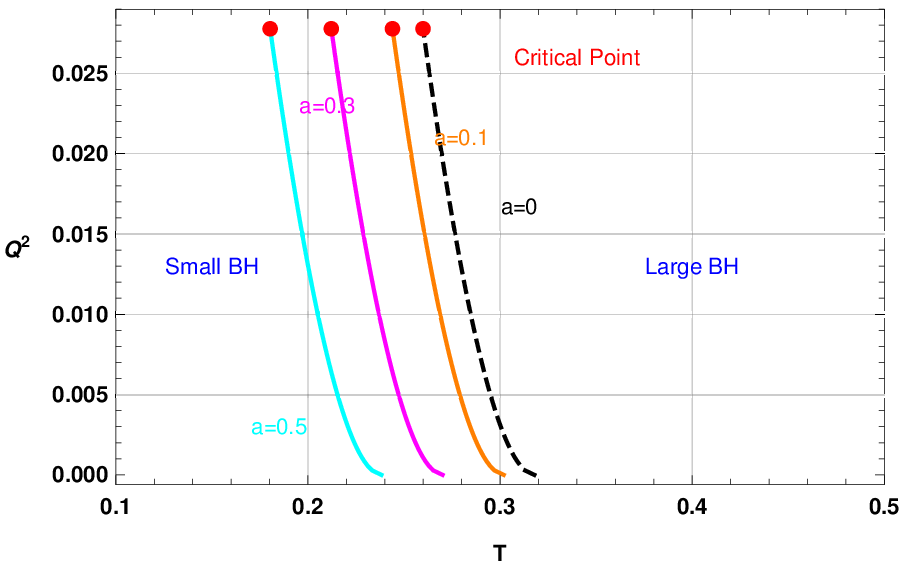} 
\end{tabbing}
\vspace*{-.2cm} \caption{\footnotesize {\bf Left}: The transition line of small-large black hole in $Q^2-T$ diagram. The critical point  marks the end of the transition line.  {\bf Right}: The effect of the quintessence parameter $a$ on the coexistence line.}\label{fig3}
\end{figure}
\end{center}

From Fig.\ref{fig2}, we  see that the Gibbs free energy develops a "swallow tail'' for
$T>T_c$, which is a typical feature in a first-order phase transition between small and large black hole. Below the critical temperature $T_c$, this "swallow tail"  disappears. The charge square and the temperature of the black hole are constants during the phase transition. Using the Maxwell equal-area construction and  the Gibbs free energy we reveal in Fig.\ref{fig3} the line where both phases of small and large black hole coexist.

The coexistence line terminates at a critical point where the phase transition transform becomes of second order. We can also see from the right panel  Fig.\ref{fig3} that the normalisation factor $a$ reduces the small black hole region, in addition the extremal large black hole is absent.  

In the subsequent analysis, we describe the critical behavior near the critical point in this alternative extended phase space. For this,  we usually introduces the following critical exponents, 
\begin{equation}
 C_\Psi= |t|^{-\alpha}\;\;\;,
 \eta= |t|^{\beta}\;\;\;,
 \kappa_T= |t|^{-\gamma}\;\;\; \text{and} \;\;\;
 |Q^2-Q^2_c|= |\Psi-\Psi_c|^{\delta}.
\end{equation}

where the exponents $\alpha$ ,$\beta$, $\gamma$ and $\delta$ describe the behavior of specific heat $C_\Psi$, the order parameter $\eta$, the isothermal compressibility coefficient and the critical isotherm respectively.

Before proceeding with the calculations of the above critical exponents, it is more convenient to use instead the reduced parameters,
 \begin{equation}\label{reduced}
  \Psi_r \equiv \frac{\Psi}{\Psi_c},\quad  Q^2_r\equiv \frac{Q^2}{Q^2_c},\quad T_r\equiv \frac{T}{T_c}.
 \end{equation}

that can be expressed as,
\begin{equation}
 T_r=1+t,\quad \Psi_r=1+\psi, \;\;\;\text{and}\;\; Q^2_r=1+\varrho,
\end{equation}
where the new variable $t$, $\psi$ and $\varrho$ describe the departure from the critical point. The first step consists of deriving the heat capacity at fixed potential $\Psi$, 

\begin{equation}
 C_\Psi=T\left.\left(\frac{\partial T}{\partial S}\right)^{-1}\right|_\Psi=0.
\end{equation}
Hence the critical exposant associated with the heat capacity is null $\alpha=0$. For the second critical exponent, we use the equation of state in the reduced variable, then  Eq.\ref{state} becomes

\begin{equation}
	Q^2_r=T_r \left(\frac{2 \sqrt{6} a l}{\Psi_r^3}-\frac{8}{\Psi_r^3}\right)+\frac{a\cdot 2^{\left(\frac{3 \omega_q }{2}+\frac{3}{2}\right)}\cdot 3^{\left(\frac{3 \omega_q
			}{2}+\frac{5}{2}\right)} \omega_q  \left(\frac{l}{\Psi_r}\right)^{-3 \omega_q }}{l \Psi_r}+\frac{3}{\Psi_r^4}+\frac{6}{\Psi_r^2}.
	\end{equation}

By expanding this equation near the critical point and using Eq. $24$, we get 

\begin{equation}\label{20}
 \varrho= -8 t+2 \sqrt{6} a l t+t \psi  \left(24-6 \sqrt{6} a l\right)-4 \psi ^3  +\mathcal{O}(t\psi^2,\phi^3).
\end{equation}

Then by differentiating Eq.(\ref{20}) with respect to $\psi$ and $t$, and  making use of the Maxwell equal area law,  we obtain
\begin{eqnarray}\label{21}\nonumber
\varrho &=& -8 t+2 \sqrt{6} a l t+t \psi_l  \left(24-6 \sqrt{6} a l\right)-4 \psi_l ^3= -8 t+2 \sqrt{6} a l t+t \psi_s  \left(24-6 \sqrt{6} a l\right)-4 \psi_l ^3,\\
0 &=& \Psi_c\int_{\psi_l}^{\psi_s} \left(t \left(24-6 \sqrt{6} a l\right)-12 \psi ^2\right)d\psi,
\end{eqnarray}
where the indices $l$ and $s$ refer to large and small black hole phase respectively. A non trivial solution of Eq.(\ref{21}) is then derived, 
\begin{equation}
 \psi_s=-\psi_l=\frac{\sqrt{t \left(4-\sqrt{6} a l\right)}}{\sqrt{2}}.
\end{equation}
So in the vicinity of the critical point, we can find the order parameter as, 
\begin{equation}
 |\psi_s-\psi_l|=2\psi_s= \frac{\sqrt{ \left(4-\sqrt{6} a l\right)}}{\sqrt{2}}t^{\frac{1}{2}}\;\; \Rightarrow \beta= 1/2.
\end{equation}

As to the isothermal compressibility coefficient $\kappa_T$ and the value of the critical exponent $\gamma$, we obtain
\begin{equation}
 \kappa_T=\left.\frac{\partial \Psi}{ \partial Q^2}\right|_T\propto \frac{\Psi_c}{(24-6 \sqrt{6} a l )Q^2_c t}\;\; \Rightarrow \gamma=1.
\end{equation}
For the last critical exponent $\delta$, we use Eq.(\ref{20}) in the isothermal case $t=0$ and get
\begin{equation}
 \left.\varrho\right|_{t=0}=-4\psi^3 \Rightarrow \delta=3.
\end{equation}

These critical exponents are similar to those of the Van der Waals fluid \cite{KM}, They can be attributed to the effect of the mean field theory. 

 In the next section, we will investigate the critical behavior of the black holes through the use of geometrical method.
 
\section{Geothermodynamics of charged black hole surrounded by quintessence}

Recently, the geothermodynamics approach has gained more attention as an efficient tool to describe geometrically the behavior of thermodynamic systems. This approach has produced consistent results to many issues related to systems such as black holes, ideal or Van der Waals gas, and cosmological models.

Here, we use the Ruppeiner metric \cite{10} to probe the effect the quintessence on the microscopical structure of such black hole,  and define this metric on the $(M,Q^2)$ space as,

\begin{equation}
 g^R_{\mu\nu}=\frac{1}{T}\frac{\partial^2 M}{\partial X^\mu\partial X^\nu},
\end{equation}
with $X^\mu=(S,Q^2)$. The invariant Ricci scalar  for any value of the state parameter of the quintessence $\omega_q$, with  $l$ set to unity, is given by

\begin{equation}
R^R=\frac{8 \Psi ^2 \left(2 l^2 \Psi ^2 \left(3 a 8^{\omega_{q} } \omega_{q}  (3 \omega_{q} -1) \Psi ^{3
		\omega_{q} +1}-1\right)-3\right)}{\pi  \left(4 l^2 \Psi ^2 \left(2 \Psi  \left(2 Q^2 \Psi
	-3 a 8^{\omega_{q} } \omega_{q}  \Psi ^{3 \omega_{q} }\right)-1\right)-3\right)}.
\end{equation}

By considering a  special case with $\omega_q=-2/3$, and using the reduced coordinates, we can recast $R^R$ into the form,

\begin{equation}\label{rrw}
 R^R\left.\right|_{\omega=-2/3}=\frac{18 \Psi_r^2 \left(\sqrt{6} a \Psi_r-2 \left(\Psi_r^2+1\right)\right)}{\pi  \left(2 \sqrt{6} a \Psi_r+Q_r^2
   \Psi_r^4-6 \Psi_r^2-3\right)}.
\end{equation}

We know that the sign of the $R^R$ can be interpreted in terms of intermolecular interaction in thermodynamical system. The positivity/negativity  refers to the repulsive/attractive
interaction  between the constituent of the system \cite{c12,c13}. Note here that the interactions are absent in the system for a null Ricci scalar $R^R$ \cite{c14} as in the classical ideal gas. 
Therefore, our analysis will focus on the sign  of the Ricci scalar $R^R$ shown in Eq.\ref{rrw}. 

    \begin{center}
\begin{figure}[!ht]
\centering
\includegraphics[scale=1]{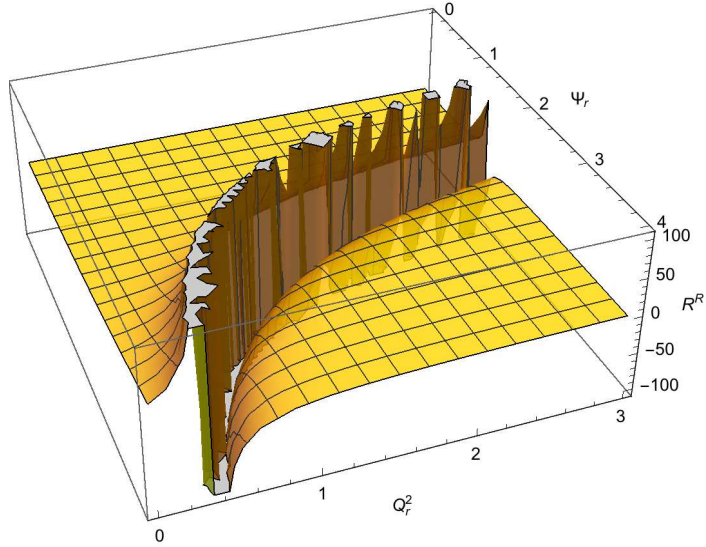} 
\vspace*{-.2cm} \caption{\footnotesize The Ruppeiner scalar curvature versus the reduced variables $\Psi_r$  and $Q^2_r$, with $a=.5$}\label{figrr}
\end{figure}
\end{center}

As can be seen from Figure.\ref{figrr}, which illustrates   $R^R$ as a function the reduced variables $\Psi_r$  and $Q^2_r$, $R^R$ does not show a monotonic sign.  Since the numerator of $R^R$ is always negative, the sign of the scalar curvature is hence controlled by the opposite  sign  of  the denominator. The latter has a root  at  
 $\Psi_0$ given by,
\begin{eqnarray}\nonumber
\Psi_0&=& \frac{1}{2} \left(\sqrt{\frac{4 \sqrt{6} a}{Q_r^2
   \sqrt{\frac{4}{Q_r^2}-\frac{2 \sqrt[3]{2}
   \left(Q_r^2-1\right)}{Q_r^4 X}+2^{2/3} X}}-\frac{2
   \sqrt[3]{2}}{Q_r^4 X}+\frac{\frac{2
   \sqrt[3]{2}}{X}+8}{Q_r^2}-2^{2/3}
   X}\right.\\
  &-&  \left.\sqrt{\frac{4}{Q_r^2}-\frac{2 \sqrt[3]{2}
   \left(Q_r^2-1\right)}{Q_r^4 X}+2^{2/3} X}\right).
\end{eqnarray}

The Ruppeiner curvature is singular at this point with $X$ expressed as,
\begin{equation}
X=\sqrt[3]{\frac{\text{Qr} \left(3 \left(a^2-2\right) \text{Qr}+\sqrt{9 a^4
   \text{Qr}^2-12 a^2 \left(3 \text{Qr}^2+1\right)+4
   \left(\text{Qr}^2+3\right)^2}\right)-2}{\text{Qr}^6}}.
\end{equation}

From the relation between the sign of $R^R$ and the nature of the intermolecular interaction, we can summarize our results about the sign of $R^R$ and the temperature  $T$  in the following table \ref{tab}.
 
\begin{table}[h!t]
\begin{center}\begin{tabular}{l|l|l|l|l|l|l}\cline{2-3}
 \hline\centering
              &$\Psi<\Psi_0$ &   $\Psi>\Psi_0$     \\ \hline\hline
$R^R$  &    \centering{+}    &-       \\ \hline\hline  
$T_0$  &    +    &-       \\ \hline\hline  
Validity  &     Allowed   &  Forbidden  \\\hline
 \end{tabular}
\end{center}
\caption{Sign of $R^R$ and its domain of validity with respect to the temperature sign.}\label{tab}
\end{table}

\begin{center}
	\begin{figure}[!ht]
		\centering
		\begin{tabbing}
			\hspace{9cm}\=\kill
			\includegraphics[width=8cm,height=7cm]{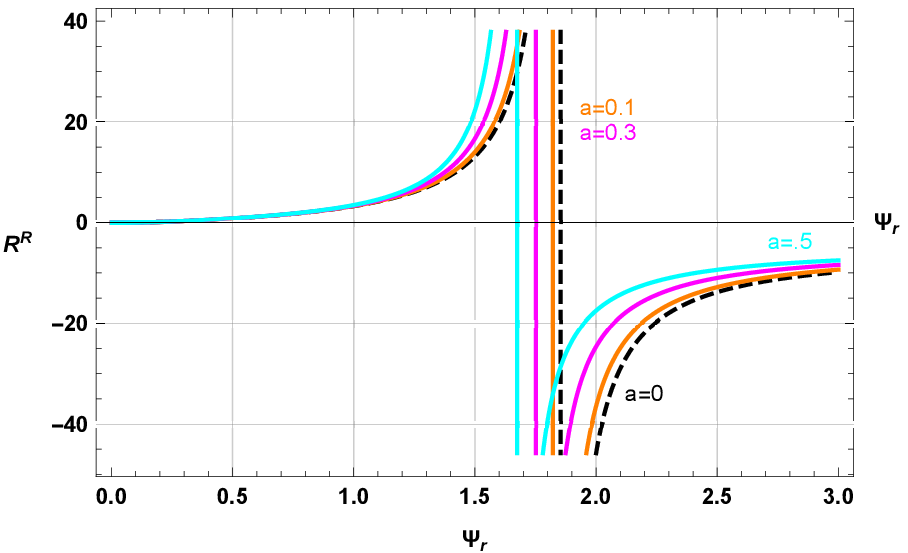}\> \includegraphics[width=7cm,height=7cm]{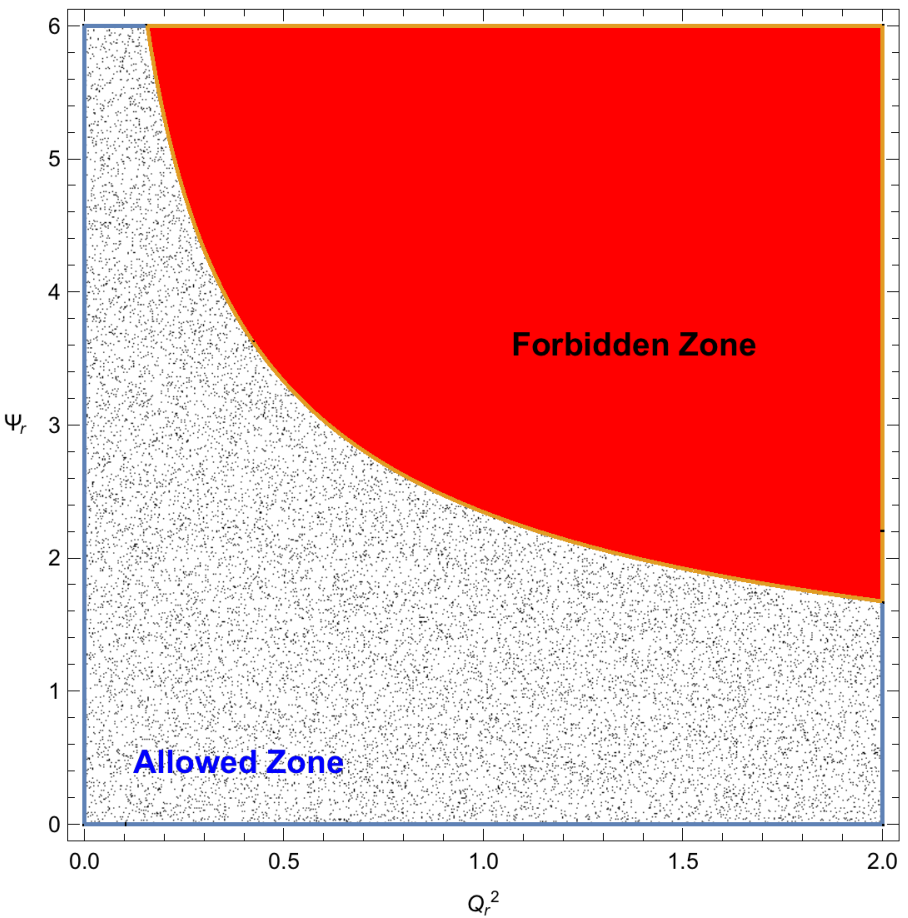} 
		\end{tabbing}
		\vspace*{-.2cm} \caption{\footnotesize {\bf Left:}  The Ruppeiner scalar curvature in function of $\Psi_r$ for different values of $a$, here we have set $Q^2_r=2$. {\bf Right:} Region plot show the region allowed physically in the extended phase space}\label{fig5}
	\end{figure}
\end{center}

The requirement of a positive absolute  black hole temperature constrains the phase space to be divided to  allowed and forbidden regions 
 of $\Psi$ and charge square as shown  in the right panel  of Fig. \ref{fig5}.
  
  Form the left panel of Fig.\ref{fig5} we can see that for $0 \leq \Psi_r\leq \Psi_0$, the Ruppeiner  curvature is positive which is equivalent to a repulsive intermolecular interaction.  This region can be divided in two parts with respect to $\Psi_r=1$:
\begin{itemize}  
  \item When $1\leq \Psi_r $  and tends to $\Psi_0$, $R^R$ becomes very large and divergent which could be  interpreted by a huge repulsive interaction. This behavior is consistant with Fermi gas near null temperature, $T\approx 0$, \cite{Dolan:2014woa}, when the Fermi exclusion principe dominates the thermodynamic behavior of the system with strong degenerate pressure. 
  \item For the second part,  characterized by $0 \leq \Psi_r\leq 1$, we  observe a small value of $R^R$ for $\Psi_r$ approaching zero which signals a rather weak repulsive interaction in the stable large black hole phase. 
\end{itemize}

Another remark  is in order at this stage: it is worth to point out the possible connection between the repulsive intermolecular interaction and the small/large black hole phase transition,  since the transition from small to large black holes is governed by the repulsive nature of the interacting constituents which tend to expand the black hole, which corroborates the result obtained in \cite{micro}.

  At last, we briefly comment on the effect  the quintessence parameter $a$. From the left panel of Figure \ref{fig5},  we can see that when $a$ increases  the values of the $\Psi_0$ decrease reducing  the allowed region, a behavior which might originate from  the interaction of the quintessence field and the constituents of the black hole . 
  
\section{Conclusion}

By proposing an alternative extended phase space defined by the square of the charge $Q^2$ and its thermodynamical conjugate quantity where the cosmological constant is kept constant, we have studied the small/large black holes phase transition of the Reissner-Nordstrom-AdS black hole surrounded by quintessence.  We have shown that such black hole exhibits a critical behavior similar to Van der Waals one, a feature characterized by the critical point, critical exponents, and a universal constant.

We have also investigated the microscopic  structure  of the charged AdS black holes surrounded by quintessence using the Ruppeiner thermodynamic geometry. This allow connection between the sign of the Ruppeiner scalar curvature, the nature of intermolecular interaction and the thermodynamical phase transition. The latter is caused by a strongly repulsive interaction amongst the constituents. Finally, we have discussed the effect of the quintessence parameters on this phase transition.


\begin{thebibliography}{99}

\bibitem{2} 
  S.~W.~Hawking and D.~N.~Page,
  {\em Thermodynamics of Black Holes in anti-De Sitter Space},
  Commun.\ Math.\ Phys.\  {\bf 87}, 577 (1983).
%  doi:10.1007/BF01208266

%\cite{Sekiwa:2006qj}
\bibitem{12} 
  Y.~Sekiwa,
  {\em Thermodynamics of de Sitter black holes: Thermal cosmological constant},
  Phys.\ Rev.\ D {\bf 73}, 084009 (2006)
  %doi:10.1103/PhysRevD.73.084009
  [hep-th/0602269].

\bibitem{13} 
  C.~Teitelboim,
  {\em The Cosmological Constant As A Thermodynamic Black Hole Parameter},
  Phys.\ Lett.\ B {\bf 158}, 293 (1985).
  %doi:10.1016/0370-2693(85)91186-4

\bibitem{14} 
  B.~P.~Dolan,
  {\em The cosmological constant and the black hole equation of state},
  Class.\ Quant.\ Grav.\  {\bf 28}, 125020 (2011)
  %doi:10.1088/0264-9381/28/12/125020
  [arXiv:1008.5023 [gr-qc]].
  
   \bibitem{KM} D.~Kubiznak and  R.~B.~Mann, {\em P-V criticality of charged AdS black holes}, J. High Energy Phys.
 {\bf 1207} (2012) 033.

% \bibitem{chin1} C.~Song-Bai, L.~Xiao-Fang, L.~Chang-Qing, {\em $P$-$V$ Criticality of an AdS Black Hole in f(R) Gravity},
%  Chin. Phys. Lett.,  {\bf 30} (2013) 060401.

\bibitem{our} A.~Belhaj, M.~Chabab, H.~El Moumni, M.~B.~Sedra, {\em
On Thermodynamics of AdS Black Holes in Arbitrary Dimensions}, Chin. Phys.Lett.
{\bf 29} 10 (2012)100401.

\bibitem{our1} A.~Belhaj, M.~Chabab, H.~El Moumni, L.~Medari, M.~B.~Sedra, {\em
The Thermodynamical Behaviors of Kerr-Newman AdS Black Holes}, Chin. Phys. Lett.
{\bf 30} (2013) 090402.

\bibitem{our2}  A.~Belhaj, M.~Chabab, H.~El Moumni, K.~Masmar and M.~B.~Sedra,
  {\em Critical Behaviors of 3D Black Holes with a Scalar Hair},
  Int.\ J.\ Geom.\ Meth.\ Mod.\ Phys.\  {\bf 12}, no. 02, 1550017 (2014)
  %doi:10.1142/S0219887815500176
  [arXiv:1306.2518 [hep-th]].
  
 \bibitem{our3}
 A.~Belhaj, M.~Chabab, H.~El moumni, K.~Masmar and M.~B.~Sedra,
  {\em Maxwell`s equal-area law for Gauss-Bonnet-Anti-de Sitter black holes},  Eur.\ Phys.\ J.\ C {\bf 75} (2015) 2,  71
  [arXiv:1412.2162 [hep-th]]. 
  
  
  \bibitem{our4} 
  A.~Belhaj, M.~Chabab, H.~El Moumni, K.~Masmar, M.~B.~Sedra and A.~Segui,
  {\em On Heat Properties of AdS Black Holes in Higher Dimensions},
  JHEP {\bf 1505}, 149 (2015)
  %doi:10.1007/JHEP05(2015)149
  [arXiv:1503.07308 [hep-th]].
  
  \bibitem{our5} 
  A.~Belhaj, M.~Chabab, H.~El Moumni, K.~Masmar and M.~B.~Sedra,
  {\em On Thermodynamics of AdS Black Holes in M-Theory},
  Eur.\ Phys.\ J.\ C {\bf 76}, no. 2, 73 (2016)
  %doi:10.1140/epjc/s10052-016-3928-9
  [arXiv:1509.02196 [hep-th]].
  
  \bibitem{our6} 
 M.~Chabab, H.~El Moumni and K.~Masmar,
  {\em On thermodynamics of charged AdS black holes in extended phases space via M2-branes background},
  Eur.\ Phys.\ J.\ C {\bf 76}, no. 6, 304 (2016)
  %doi:10.1140/epjc/s10052-016-4155-0
  [arXiv:1512.07832 [hep-th]].
  
  
  \bibitem{our7} 
  M.~Chabab, H.~El Moumni, S.~Iraoui and K.~Masmar,
  {\em Behavior of Quasinormal Modes and high dimension RN-AdS Black Hole phase transition}, Eur.\ Phys.\ J.\ C {\bf 76}, no. 12, 676 (2016),
  arXiv:1606.08524 [hep-th].
  
 %\cite{ElMoumni:2016eqh}
\bibitem{ElMoumni:2016eqh} 
  H.~El Moumni,
  {\em Phase Transition of AdS Black Holes with Non Linear Source in the Holographic Framework},
  Int.\ J.\ Theor.\ Phys.\  {\bf 56}, no. 2, 554 (2017).
  %doi:10.1007/s10773-016-3197-2
  %%CITATION = doi:10.1007/s10773-016-3197-2;%%
  %1 citations counted in INSPIRE as of 14 Apr 2017
  
  
%\cite{Kiselev:2002dx}
\bibitem{zh38} 
  V.~V.~Kiselev,
  {\em Quintessence and black holes},
  Class.\ Quant.\ Grav.\  {\bf 20}, 1187 (2003)
  %doi:10.1088/0264-9381/20/6/310
  [gr-qc/0210040].
  %%CITATION = doi:10.1088/0264-9381/20/6/310;%%
  %89 citations counted in INSPIRE as of 13 Aug 2016

  %\cite{Chen:2005qh}
\bibitem{24} 
  S.~b.~Chen and J.~l.~Jing,
  {\em Quasinormal modes of a black hole surrounded by quintessence},
  Class.\ Quant.\ Grav.\  {\bf 22}, 4651 (2005)
  %doi:10.1088/0264-9381/22/21/011
  [gr-qc/0511085].
  
  
    %\cite{Spallucci:2013jja}
\bibitem{Spallucci:2013jja} 
  E.~Spallucci and A.~Smailagic,
  {\em Maxwell's equal area law and the Hawking-Page phase transition},
  J.\ Grav.\  {\bf 2013}, 525696 (2013)
  %doi:10.1155/2013/525696
  [arXiv:1310.2186 [hep-th]].
  %%CITATION = doi:10.1155/2013/525696;%%
  
  
 %\cite{Zhang:2006ij}
\bibitem{25} 
  Y.~Zhang and Y.~X.~Gui,
  {\em Quasinormal modes of a Schwarzschild black hole surrounded by quintessence},
  Class.\ Quant.\ Grav.\  {\bf 23}, 6141 (2006)
  %doi:10.1088/0264-9381/23/22/004
  [gr-qc/0612009]. 
  
  \bibitem{26} 
  S.~Chen, B.~Wang and R.~Su,
  {\em Hawking radiation in a $d$-dimensional static spherically-symmetric black Hole surrounded by quintessence},
  Phys.\ Rev.\ D {\bf 77}, 124011 (2008)
  %doi:10.1103/PhysRevD.77.124011
  [arXiv:0801.2053 [gr-qc]].
  \bibitem{27}
    Y.~H.~Wei and Z.~H.~Chu,
  {\em  Thermodynamic properties of a Reissner-Nordstroem quintessence black hole},
  Chin.\ Phys.\ Lett.\  {\bf 28}, 100403 (2011).
 % doi:10.1088/0256-307X/28/10/100403
  
  
  \bibitem{28} 
  M.~Azreg-Aïnou and M.~E.~Rodrigues,
  {\em Thermodynamical, geometrical and Poincaré methods for charged black holes in presence of quintessence},
  JHEP {\bf 1309}, 146 (2013)
 % doi:10.1007/JHEP09(2013)146
  [arXiv:1211.5909 [gr-qc]].
  %%CITATION = doi:10.1007/JHEP09(2013)146;%%
  %12 citations counted in INSPIRE as of 14 Apr 2017  
  
  
  %\cite{Li:2014ixn}
\bibitem{29} 
  G.~Q.~Li,
  {\em Effects of dark energy on P-V criticality of charged AdS black holes},
  Phys.\ Lett.\ B {\bf 735}, 256 (2014)
 % doi:10.1016/j.physletb.2014.06.047
  [arXiv:1407.0011 [gr-qc]].
  
  %\cite{Liu:2017baz}
\bibitem{machine} 
  H.~Liu and X.~H.~Meng,
  {\em Effects of dark energy on the efficiency of charged AdS black holes as heat engine},
  arXiv:1704.04363 [hep-th].
  
  %\cite{Ma:2016arz}
\bibitem{ma} 
  M.~S.~Ma, R.~Zhao and Y.~Q.~Ma,
  {\em Thermodynamic stability of black holes surrounded by quintessence},
  arXiv:1606.06070 [gr-qc].
  
  
    %\cite{Zeng:2015wtt}
\bibitem{holo} 
  X.~X.~Zeng and L.~F.~Li,
  {\em Van der Waals phase transition in the framework of holography},
 Phys.\ Lett.\ B {\bf 764}, 100 (2017)
  %doi:10.1016/j.physletb.2016.11.017
  [arXiv:1512.08855 [hep-th]].
  
  
    %\cite{Dehyadegari:2016nkd}
\bibitem{micro} 
  A.~Dehyadegari, A.~Sheykhi and A.~Montakhab,
  {\em Microscopic properties of black holes via an alternative extended phase space},
 Phys.\ Lett.\ B {\bf 768}, 235 (2017)
  %doi:10.1016/j.physletb.2017.02.064
  [arXiv:1607.05333 [gr-qc]].
  %%CITATION = doi:10.1016/j.physletb.2017.02.064;%%
  %1 citations counted in INSPIRE as of 14 Apr 2017
  
  
    %\cite{Sahay:2010tx}
\bibitem{10} 
  A.~Sahay, T.~Sarkar and G.~Sengupta,
  {\em On the Thermodynamic Geometry and Critical Phenomena of AdS Black Holes},
  JHEP {\bf 1007}, 082 (2010)
  %doi:10.1007/JHEP07(2010)082
  [arXiv:1004.1625 [hep-th]].
  %%CITATION = doi:10.1007/JHEP07(2010)082;%%
  %36 citations counted in INSPIRE as of 09 Mar 2017
  
  
  %\cite{Wang:2002nq}
%\bibitem{30} 
%  B.~B.~Wang and C.~G.~Huang,
%  {\em Thermodynamics of Reissner-Nordstrom-De Sitter black hole in York's formalism},
%  Class.\ Quant.\ Grav.\  {\bf 19}, 2491 (2002).
  
  
  %\cite{Wei:2015iwa}
\bibitem{c12} 
  S.~W.~Wei and Y.~X.~Liu,
  {\em Insight into the Microscopic Structure of an AdS Black Hole from a Thermodynamical Phase Transition},
  Phys.\ Rev.\ Lett.\  {\bf 115}, no. 11, 111302 (2015)
  Erratum: [Phys.\ Rev.\ Lett.\  {\bf 116}, no. 16, 169903 (2016)]
 % doi:10.1103/PhysRevLett.116.169903, 10.1103/PhysRevLett.115.111302
  [arXiv:1502.00386 [gr-qc]].
  
  
  %\cite{Zangeneh:2016snh}
\bibitem{c13} 
  M.~K.~Zangeneh, A.~Dehyadegari and A.~Sheykhi,
  {\em Comment on "Insight into the Microscopic Structure of an AdS Black Hole from a Thermodynamical Phase Transition},
  arXiv:1602.03711 [hep-th].
  %%CITATION = ARXIV:1602.03711;%%
  
  
 \bibitem{c14} 
  G. Ruppeiner. 1979. Phys.Rev.{\bf A 20}1608. 
  
  
  %\cite{Dolan:2014woa}
\bibitem{Dolan:2014woa} 
  B.~P.~Dolan,
  {\em Black holes, the van der Waals gas, compressibility and the speed of sound},
  Fortsch.\ Phys.\  {\bf 62}, 892 (2014).
  %doi:10.1002/prop.201400011
  %%CITATION = doi:10.1002/prop.201400011;%%
  %3 citations counted in INSPIRE as of 07 Mar 2017
  
  

  
\end{thebibliography}
\end{document}